\documentclass [epjST] {svjour}
\usepackage{graphicx}
\usepackage[utf8]{inputenc}
\usepackage{gensymb}
\usepackage{cite}
\begin{document}
\title{Hollow Core, Whispering Gallery Resonator Sensors}
\author{Jonathan M. Ward\thanks{\email{jonathan.ward@oist.jp}} \and Nitesh Dhasmana \and S\'{i}le Nic Chormaic}
\institute{Light-Matter Interactions Unit, OIST Graduate University, 1919-1 Tancha,\\Onna-son, Okinawa 904-0495 Japan }
\abstract{A review of hollow core whispering gallery resonators (WGRs) is given. After a short introduction to the topic of whispering gallery resonators we provide a description of whispering gallery modes in hollow or liquid core WGRs. Next, whispering gallery mode (WGM) sensing mechanisms are outlined and some fabrication methods for microbubbles, microcapillaries and other tubular WGM devices are discussed.   We then focus on the most common applications of hollow core WGRs, namely refractive index and temperature sensing, gas sensing, force sensing, biosensing, and lasing. The review highlights some of the key papers in this field and gives the reader a general overview of the current state-of-the-art.
} %end of abstract
\maketitle
\section{Introduction}
\label{intro}
Whispering gallery resonators (WGRs) have come of age in terms of their application to optical sensing and, in the last few decades, various novel geometries have appeared, each with their own specific advantages and disadvantages. The theory of scattering by small dielectric particles was developed by Mie \cite{1} and Debye \cite{2} at the start of the $20^{th}$ century. The suggestion that spherical dielectric structures could act as high frequency resonators and support high quality optical modes was first published in 1939 by Richtmyer \cite{3}. The first optical WGR experiments were carried out in the  1960's by Garrett et al. \cite{4}; the whispering gallery modes (WGMs) were created in doped crystalline microspheres with diameters of a few millimetres. The Sm$^{+}$ doped microspheres, which were able to sustain lasing, were excited by focussing a pump laser onto the edge of the sphere. Ashkin and Dziedzic reported on the observation of WGM resonances in the scattering spectra of liquid droplets trapped in an optical tweezers \cite{5}, and the topic was further studied by Ch\'{y}lek et al. \cite{6}.  Subsequently, liquid droplets containing dye molecules, which acted as a gain medium, were used as WGR lasers in the early 1980's \cite{7}. From this point the interest in WGRs increased significantly and, after the development of robust coupling using evanescent field couplers, such as prisms \cite{8} and tapered optical fibres \cite{9,21a},  the realisation of the WGR as a useful optical device became apparent.   Many passive and active optical components such as  filters \cite{10}, selective feedback devices \cite{11}, incoherent light sources \cite{11a}, WGR lasers ~\cite{7,12,12a,13}, and delay lines ~\cite{14,15,16} were developed.  At the same time, the number of materials and geometries being used increased, with polymers and crystalline materials being added \cite{19a} to the already numerous types of glass out of which WGRs were fabricated. These materials have been molded into a wide variety of forms from disks \cite{17} and toroids \cite{18} to goblets \cite{19}, peanuts \cite{20}, bottles \cite{21,21a,22} and bubbles \cite{23,23a}. Today, WGRs can be found in experiments in nonlinear optics \cite{24,25}, cavity optomechanics \cite{26,27}, cavity QED \cite{28}, and chaos \cite{29,30}.  The WGR sensor was also developed during this time and further progress in the 2000's led to WGRs being used to measure many physical phenomena, such as refractive index, temperature, force/pressure, as well as for chemical and biosensing. In this paper we will focus on the WGM sensor in the form of the hollow or liquid core optical resonator (LCOR).

\section{EM Modes in Hollow WGM Resonators}
%\label{sec:1}
% \cite{RefJ}
WGRs are typically solid, micron-scale, dielectric structures that trap light by continuous total internal reflection. The classic example is a solid glass microsphere \cite{30a,31} whereby light is coupled into the microsphere using an evanescent field coupler such as a tapered optical fibre, integrated optical waveguide, or a prism \cite{32}. An evanescent field occurs wherever light undergoes total internal reflection, i.e. at the boundary between two media.  The light extends into the lower index material with an intensity that decays exponentially from the interface.  When the microsphere is placed into the evanescent field of an optical waveguide the light can tunnel from the waveguide into the microsphere where it propagates as a WGM. Just like in the waveguide, the WGM in the microsphere creates an evanescent field extending from the surface of the sphere \cite{32}. This evanescent tail allows the modes to interact with the environment external to the microsphere.   A schematic of a WGM in a microsphere is shown in Fig. 1(a) along with the spherical coordinate axes. Hollow WGRs, such as microbubbles or capillaries, are - as their name implies - simply, thin shells surrounding a hollow volume. A cross-sectional schematic of a hollow microsphere WGR and photographs of actual hollow microspheres are shown in Figs. 1(b) and 1(c).
\begin{figure}
{
\includegraphics[scale=.25,origin=r]{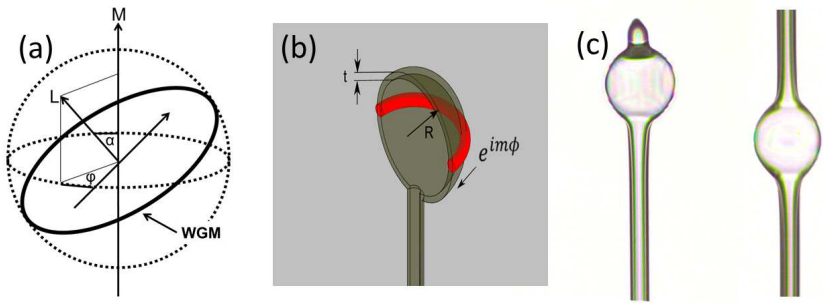} }
\caption{(a) Schematic of a WGM in a sphere showing the spherical polar coordinates. (b) Cross-sectional schematic of a hollow microsphere, the red band depicts the WGM. (c) Images of microbubble WGRs with a single input port and double input ports; both bubbles have diameter around 60 $\mu$m and wall thickness around 1 $\mu$m.}       % Give a unique label
\end{figure}

%\begin{figure}
%% Use the relevant command for your figure-insertion program
%% to insert the figure file.
%% For example, with the option graphics use
%\centering\resizebox{1\columnwidth}{!}
%{
% \centering \includegraphics{} }
%\caption{(a) Schematic of a WGM in a sphere showing the spherical polar coordinates. (b) Cross sectional schematic of a hollow microsphere, the red band depicts the WGM. (c) images of microbubble WGRs with a single input port and double input ports, both bubbles have diameters around 60 $\mu$m and wall thicknesses around 1 $\mu$m.}
%% Give a unique label
%\end{figure}

Given that the physical structure of the microbubble \cite{23,23a} is similar to that of the microsphere one would expect the distribution of the WGMs to also be similar, and indeed they are, at least in terms of the azimuthal and polar layout. For thick walled microbubbles (where the thickness, $t\gg\lambda$, the propagating wavelength) the radial characteristics are also similar.  However, this changes when the wall becomes thin (i.e. $t\approx\lambda$) and/or when the core of the microbubble is filled with some fluid. In the following discussion, let us consider the WGMs in a hollow microsphere. We will assume a spherically symmetric structure in the form of a hollow sphere, with inner and outer radii $R_{1}$ and $R_{2}$, respectively, and with refractive index $n_{2}$. The electromagnetic (EM) waves inside a microbubble can be described by solving the Helmholtz equation in spherical polar coordinates \cite{33}.  The analytical form of the electromagnetic field can be obtained for the polar and azimuthal components in the same fashion as for a microsphere. The radial part is given by \cite{34}:

\begin{equation}
%%% remove comment delimiter ('%') and select language if required
%\selectlanguage{spanish}
E_r=\left\{ \begin{array}{c}
\ \ \ \ \ \ \ Aj_m\left(k^{\left(m,l\right)}n_1r\right),\ \ \ \ \ \ \ \ \ \ \ \ \ \ \ \ \ r\le R_1 \\
Bj_m\left(k^{\left(m,l\right)}n_2r\right)+Ch^{(1)}_m\left(k^{\left(m,l\right)}n_2r\right),\ \ \ R_1<r\le R_2 \\
\ \ \ \ \ \ \  Dh^{(1)}_m\left(k^{\left(m,l\right)}n_3r\right),\ \ \ \ \ \ \ \ \ \ \ \ \ \ \  r>R_2 \end{array}
\right.
\end{equation}

where \emph{A, B, C, D} are unknowns which can be determined by applying electromagnetic boundary conditions, $j_m$, $h^{(1)}_m$ are spherical Bessel  and  Hankel functions of the first kind, respectively, with order \emph{m}, $k^{(m,l)}$ is the amplitude of the resonant wave vector component in the azimuthal direction labelled by the azimuthal index \emph{m} and the radial index \emph{l}, and  $n_{1}, n_{3}$ are the refractive indices of the medium inside and outside the microbubble resonator, respectively. The boundary conditions state that the tangential components of the electric and magnetic field must be continuous at the surface. The modes of WGM structures can also be determined using commercially available software packages such as COMSOL, which relies on the Finite Element Method (FEM). If spherical symmetry is assumed, the method by Oxborrow \cite{35} (see also \cite{36}) can be used to considerably reduce the computational time and memory.  Figure 2 shows an example of such a solution for an acetone-filled microbubble with a diameter of 50 $\mu$m and wall thickness of 1 $\mu$m. The model used perfectly matched layers, and enabled us to determine the \emph{Q} factors of the modes. The corresponding radial profiles are also shown in Fig. 2;  these plots highlight the fact that the modes have the usual evanescent field on the outside, but also an evanescent field that extends into the liquid core.  The extent to which the mode interacts with the fluid in the core depends on the wall thickness and the mode order, with higher order radial modes having a larger overlap with the core fluid. This is illustrated in Fig. 3 for a 50 $\mu$m diameter bubble.  The bubble was filled with water and the WGM wavelength was 1550 nm. The percentage of energy in the core for the first three radial modes for a shell thickness varying from 300 nm to 3 $\mu$m was calculated for fixed diameter microbubbles.
\begin{figure}
%\resizebox{.5\columnwidth}{!}
{
\includegraphics[origin=c,scale=.25]{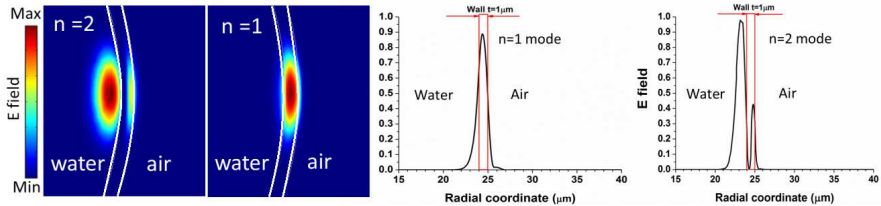} }
\caption{The electric field in a microbubble for the radial and polar directions with the corresponding radial profiles of the modes. Reproduced from \cite{37} with permissions.}       % Give a unique label
\end{figure}

The estimated percentage of the WGM's EM field in the core was found by integrating the EM field intensity in the core and shell separately. When a significant portion of a mode is in the liquid core of the cavity that mode can be considered to be in the quasi-droplet regime, i.e. the mode is similar to that of a WGM in a pure liquid droplet WGR. The onset of the quasi-droplet regime occurs for a different wall thickness for different modes and the quasi-droplet regime can be defined in terms of the mode's effective index, percentage of light in the core, or the position of the field maximum relative to the resonator's boundary \cite{37}.

\begin{figure}
%\resizebox{.5\columnwidth}{!}
{
\includegraphics[origin=r,scale=.25]{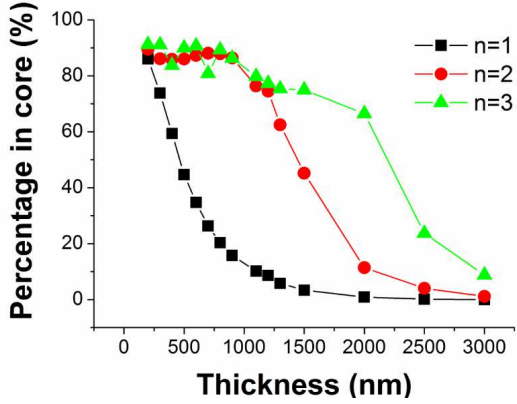} }
\caption{The percentage of light in the core as a function of the wall thickness for the first three radial modes in a microbubble with a diameter of 50 $\mu$m. Reproduced from \cite{37} with permissions.}       % Give a unique label
\end{figure}

%\begin{figure}
%\centering
%\parbox{1.2in}{\centering \includegraphics{}}
%\qquad
%\begin{minipage}{1.2in}
%\centering \includegraphics{}
%\end{minipage}%
%\caption{Here are two figures side-by-side.}
%\label{fig:1figs}%
%\end{figure}
%
%\begin{figure}
%% Use the relevant command for your figure-insertion program
%% to insert the figure file.
%% For example, with the option graphics use
%\centering\resizebox{1\columnwidth}{!}
%{
% \centering \includegraphics{} }
%\caption{Please write your figure caption here.}
%\end{figure}

\section{Sensing Mechanism}
WGM sensors are primarily based on the resonant wavelength interrogation technique. Any perturbation in the net optical path length or EM boundary conditions leads to a shift in the resonant spectrum which can be calibrated for sensing purposes. As shown before, most of the energy in a WGM remains confined, but it is not limited to being entirely within the material of the resonator. A component of the mode sits outside the resonator material in the form of an evanescent field that gives the mode an effective refractive index \cite{38}:

\begin{equation}
n_{eff}=\frac{\int{n\left(r\right){E(r)}^2dr}}{\int{{E(r)}^2dr}}
\end{equation}

In the presence of any foreign substance near the resonator, this effective index gets perturbed, leading to a change in optical path length and, in turn, the resonant spectrum. A similar effect will be seen if the physical size of the resonator changes, perhaps by adsorption or temperature changes, with the net effect given by Vollmer and Arnold in the relation \cite{39}

\begin{equation}
\frac{\triangle {\lambda }}{{\lambda }}=\frac{\triangle n_{eff}}{n_{eff}}+\frac{\triangle R}{R},
\end{equation}

where $\lambda$ is the resonant wavelength and \emph{R} is the radius of the resonator. The mechanism for sensing molecules with a size less than the penetration depth of the evanescent wave, a case which holds more importance since it covers a major subset of biomolecules, can be given more explicitly. The strong field at the surface tends to polarise the molecule/nanoparticle, which acts as a perturbation to the resonant mode of the cavity.  The amount of energy needed to polarise the nanoparticle accounts for the shift of the resonance wavelength. The interaction energy can be given as:  $\frac{1}{2}\alpha_{ex}\left|E(r_0)\right|^2$ where ${\alpha}_{ex}$ is the excess polarizability of the molecule over the medium in which the resonator is kept and $r_{0}$ is the binding site of the molecule.  This can be treated as a first order perturbation to the single photon resonant state to give the fractional change in frequency as the ratio of the interaction energy and the total electromagnetic energy of the unperturbed mode, where

\begin{equation}
\frac{\triangle \omega }{\omega }=-\frac{{\alpha }_{ex}{\left|E(r_0)\right|}^2}{2\int{\varepsilon \left|{E(r)}^2\right|\ dV}}.
\end{equation}

This is known as the reactive sensing mechanism \cite{99,100}.   The shift in the resonance wavelength can be observed by recording the transmission spectrum  whilst scanning the laser source across an interval near the resonance. The ultra-narrow line width owing to the high \emph{Q} makes the shift highly resolvable, but is, however, not the ultimate limit of the detectable shift in the resonance by any means. A fraction, \emph{F}, of the linewidth of the cavity (${\Delta\lambda}_{FWHM}$), as determined by measurement noise arising due to thermal fluctuations of the resonator and source fluctuations, can be sensed and values as low as 1/100 are achievable, implying a detectable wavelength shift of the order of fm ($10^{-15}$m) \cite{fraction,39}.   In the more pragmatic case when multiple particles attach to the resonator, the shift further depends on the average surface density of the particles, ${\sigma}_{p}$. Taking this into account, and substituting the expression for the field in the previous equation, an overall expression for the net shift is \cite{99}

\begin{equation}
\frac{\triangle \omega }{\omega }=-\frac{{\alpha }_{ex}{\sigma }_p}{{\varepsilon }_0R\left(n^2_2-n^2_1\right)},
\end{equation}

%insert figure

 where  $n_{2}$ and $n_{1}$ are the refractive indices of the sphere of radius $R$ and the surrounding medium, respectively. It is worth noting that $\Delta\omega \propto {\alpha}_{ex}$ implying that a particle with a larger molar mass is expected to cause a greater shift (${\alpha}_{ex}{\propto}m$) and hence a smaller number of such particles can result in a detectable shift. This can be quantified by introducing a term ‘limit of detection of surface density ($\sigma_{lod}$) given as
\begin{equation}
{\sigma }_{lod}=\frac{R\left(n^2_2-n^2_1\right)F}{{{(\alpha }_{ex}}/{\varepsilon_0)}Q}.
\end{equation}
%\subsection{Subsection title}
%\label{sec:2}
%as required. Don't forget to give each section
%and subsection a unique label (see Sect.~\ref{sec:1}).
%

To achieve a higher sensitivity one must choose  optimum values of \emph{R/Q} where \emph{Q} is the optical quality factor of the resonator. Also, ${\sigma}_{lod}$ reflects the feasibility of a particular single particle detection. So far, for a silica microsphere WGM resonator the limit of mass detection is set at 20 ag for a nanoparticle of radius $\sim17$ nm and can be used for viral detection \cite{101}, but not for proteins (mass $\sim$0.1 ag).

\section{Fabrication of Hollow WGM Resonators}
Micron scale glass capillaries are commercially available and can be used to make hollow WGRs. One method of fabrication requires that the capillaries be  tapered down to the required diameter, typically below 100 $\mu$m, and the remaining thick wall of the capillary is further reduced by chemical etching so that thin walls are achieved.  However, this etching stage can reduce the optical \emph{Q} \cite{34}. Microbubble or microbottle resonators are typically made by heating a small section of a glass capillary while at the same time pressurising the air inside the capillary. When the glass  softens the air pressure pushes on the walls to form a bottle or bubble shape. The capillary can be heated in the focus of a CO$_2$ laser beam and, to ensure uniform heating, the capillary can be rotated in the focus. This method was  first used by Sumetsky et al. in 2010 to make microbubbles with diameters around 300 $\mu$m and wall thicknesses of 3-4 $\mu$m \cite{23}. Chemical etching was later used to reduce the wall thickness to 500 nm, thereby making the first quasi-droplet resonator \cite{40}.

 \begin{figure}
%\resizebox{.5\columnwidth}{!}
{
\includegraphics[origin=l,scale=.3]{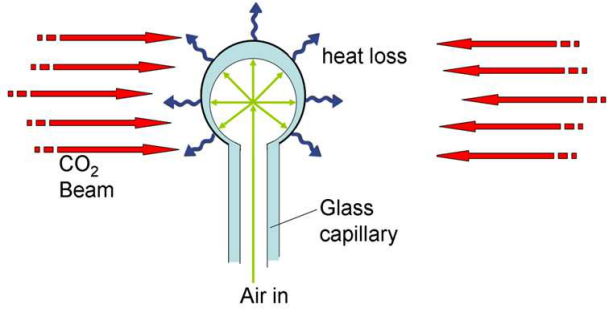} }
\caption{Schematic of the fabrication method for a microbubble WGR. A glass capillary is heated by counter propagating CO$_2$ focussed laser beams. The air inside the capillary is pressurised and pushes on the wall to form a bubble shape.}       % Give a unique label
\end{figure}

Smaller microbubbles can be made by selecting the appropriate inner and outer dimensions of the capillary and tapering the capillary diameter down to much smaller diameter e.g. $\sim$20 $\mu$m \cite{23a}.  The capillary can then be heated in focussed, counter propagating CO$_2$ beams, see Fig. 4. Because of the initially small capillary dimensions, the bubble wall thickness may become submicron during expansion. This method produces much smaller diameter bubbles ($\sim$50  $\mu$m) than other reported methods, and removes the need to rotate the capillary during heating and/or the need to chemically etch the bubble wall \cite{41,42}.   Another heat source that has been used recently is the electric arc \cite{43,44}. This method uses the electrodes from a fusion splicer to heat and soften the glass while the air inside the capillary is compressed. In \cite{43} the electrodes were rotated around the stationary capillary during the heating process. Using the fusion arc system microbubbles with diameters around 150 $\mu$m and 4 $\mu$m wall thickness were made with \emph{Q} factors of $10^7$.   These fabrication methods above could be described as “heat and expand”, but microbottle WGRs can also be made by a “heat and compress” method where the ends of the capillary are pushed together during heating. This pushing causes the heated section to bulge out, creating a bottle shape \cite{45}.

The glass capillary method is the obvious way to make hollow WGR; however, there is another method that has also been explored, i.e.  the self-rolled semiconductor. This method was  pioneered by Prinz et al. in 2000 \cite{46,47} and was extended to composite polymer/metal micro- and nanotubes \cite{48}. The fabrication relies on the stress between mismatched layers, see Fig. 5. A schematic illustration of the method is shown in Fig. 5(a).  A 1.1 nm thick InAs/GaAs bilayer is deposited on top of AlAs. The bilayer is freed by highly selective etching of the sacrificial AlAs layer in a HF etchant. The interatomic forces in the bilayer act to increase the interatomic distance in the compressed InAs layer and to decrease the interatomic distance in the tensile-stressed GaAs layer. The elastic forces $F_1$ and $F_2$ in Fig. 5(a) are oppositely directed, which tends to bend the bilayer. Under this action, the initially planar bilayer rolls up into a scroll. The rolled-up layers stick together and form a perfectly bonded tube wall. The rolled up tubes have nanometer wall thickness and diameters of a few microns.   These self-assembled tubes may have many application such as microneedles for the biomedical industry.

\begin{figure}
%\resizebox{.5\columnwidth}{!}
{
\includegraphics[origin=r,scale=.2]{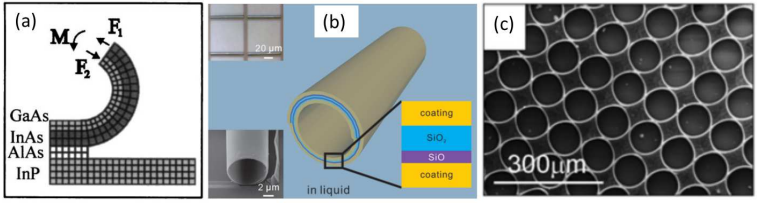} }
\caption{(a) The strain-induced self-roll mechanism for semiconductor InAs/GaAs bilayers after freeing it from the substrate. Reproduced from \cite{46} with permissions.   (b) Diagram of a rolled-up SiO/SiO$_2$ nanomembrane with additional oxide coating layers on both inner and outer surfaces.  Reproduced from \cite{49} with permissions.  (c) An array of rolled up microtoroids on a chip. Reproduced from \cite{50} with permissions.}       % Give a unique label
\end{figure}
Later, this technique was expanded to include many other materials and the sacrificial layer was replaced with a polymer which could be easily removed without damaging the bilayer. Pt,  Pd/Fe/Pd, TiO$_2$, ZnO, Al$_2$O$_3$, Si$_x$N$_y$, Si$_x$N$_y$/Ag and diamond-like carbon have been used to make tubes \cite{48}. Tubes of well-defined length and geometry were also arranged into large periodic arrays of SiO/SiO$_2$ microtubes on a Si substrate \cite{49}.  Not only can open ended tubes be made, but closed structures can also be self-assembled.  In 2007 single, and arrays of, hollow microtoroids were made (see Fig. 5(c)), although there was no attempt to excite WGMs in the devices \cite{50,51}.  On-chip microfabrication of hollow WGRs is also possible as shown in 2011 by Scholten et al.;  however, no fluidic experiments were performed \cite{52}.

\section{Hollow WGM Resonator for Pressure Sensing}
Just like solid WGRs, hollow WGRs can be used for the measurement of many physical phenomena. For example, the application of force on a WGR manifests itself as a shift of the cavity’s WGMs due to a distortion of the cavity shape or a change in refractive index as a result of stress/strain. The measurement of force has some distinct advantages in hollow WGRs over solid ones. The sensitivity of the WGM shift to an applied force is increased since the spring constant of the spherical shell is reduced compared to the solid WGR. Iopollo et al. presented calculations that compared solid and hollow PDMS microsphere WGR as force sensors \cite{54}.
One of the first demonstrations of tuning the WGMs of a hollow microcavity was given in 2010 by Sumetsky et al. \cite{55}.  In this experiment, a microbubble resonator with a diameter of 220 $\mu$m and a wall thickness of $\sim$1 $\mu$m was tuned over one FSR by stretching the microbubble in a piezo activated clamp. Stretching the capillary by 30 $\mu$m resulted in a WGM shift of 5.5 nm. No estimate of the force was given nor any discussion about the shift of the WGMs due to the dependence of the refractive index on strain. The change in the microbubble radius is a function of the applied force, $F_a$, Young's modulus of the material, $Y$, and the wall thickness, $t$, such that $\Delta$$R$$\sim0.2F_{a}/Yt$,  therefore the WGM shift could be calibrated against a known force.
The shift of the WGMs in a hollow microsphere due to changes in internal gas pressure was studied experimentally in 2011 \cite{41}. In this setup, the air inside a hollow microsphere was compressed and the resulting shift of the WGMs was measured for microbubbles with different dimensions. Using this novel method, WGM shifts from tens to hundreds of GHz were achieved with the total shift depending on the diameter and wall thickness of the microbubble. The shift rate of the WGMs from the theory of elasticity is given as \cite{41}

\begin{equation}
\frac{\triangle \lambda(p_i)}{\lambda}=\frac{2n_0b^3+12CGb^3}{4Gn_0(a^3-b^3)}p_i-\frac{n_0(a^3+b^3)+12CGa^3}{4Gn_0(a^3-b^3)}p_{0},
\end{equation}
where $a$ and $b$ denote the outer and inner radii of this shell, while $p_o$ and $p_i$ are the externally and internally applied uniform pressures. $C$ and $G$ are the elasto-optic coefficients and shear modulus of the material.  Figure 6 shows the measured WGM shift in different microbubbles; the solid lines are the theoretical fits from Eq. (7) and accurate fitting was only possible by taking into account both the change in size and the strain dependence of the refractive index.

\begin{figure}
%\resizebox{.5\columnwidth}{!}
{
\includegraphics[origin=r,scale=.3]{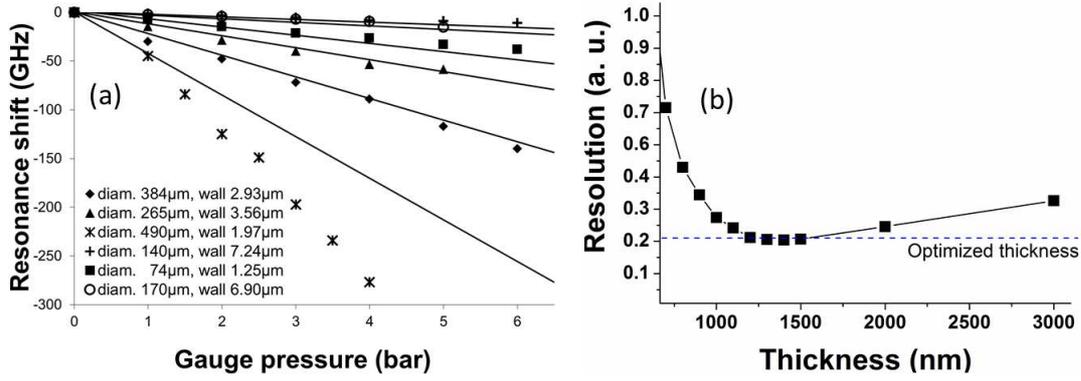} }
\caption{(a) WGM shifts as function of internal gas pressure in microbubbles of different dimensions. (b) The sensing resolution of a microbubble WGR as a function of wall thickness for an air-filled microbubble with a diameter of 50 $\mu$m. The best sensing resolution occurs at an optimal wall thickness of 1.25 $\mu$m. Reproduced from \cite{37} with permissions.}       % Give a unique label
\end{figure}

From Eq. (7) it can be derived that the total resonance shift in a microbubble is, by neglecting the constant term depending on $p_0$,  directly proportional to a geometrical parameter $\chi=b^3/(a^3-b^3)$ and, therefore, to the ratio between the whole enclosed volume and the volume of the spherical shell alone. With this relation it is possible to estimate the pressure-tuning abilities of the microbubbles directly after production without any previous measurement just from the principal diameters. With the given values for the material parameters of the capillary the theoretical proportionality factor between the slope of the resonance shift and the geometrical parameter $\chi$ can be calculated approximately as -1.1 GHz/bar. It should be noted that the work by Yang et al. \cite{37}  showed that the \emph{Q} factor of air-filled microbubble WGRs is dependent on the wall thickness, mode order and wavelength.  Therefore, there is an optimal wall thickness to give the best sensing resolution.   This is shown by the curve in Fig. 6 which, for a wavelength of 1550 nm and bubble diameter of 50 $\mu$m, shows that the optimal wall thickness for measuring the gas pressure change is around 1.25 $\mu$m \cite{37}.

\section{Temperature Sensing}
The sensitivity of WGM resonances to changes in temperature has been studied extensively in the literature over the years \cite{56}-\cite{59a}. The WGMs in silica glass microcavities experience a red shift with increasing temperature due to the positive thermo-optical coefficient and the positive thermal expansion coefficient of the glass. The ratio between these two properties depends on the material. In silica, the coefficients add to give a shift rate of $\sim$ 3 GHz/K.  The thermal shifting of WGMs in microcavities can be a help or a hindrance depending on the application. For example, the thermally induced optical bistability can make a useful optical switch or allow thermal locking of the cavity  resonance to a laser, whereas  in bio/chemical sensing the thermal drifting can be a source of noise that needs to be controlled. Hybrid microcavities, i.e. multi-layer cavities, have shown interesting thermal characteristics, whether it be improved thermal stability or increased sensitivity. The thermal behaviour of the WGMs in hollow WGRs can be controlled by selection of the core material and/or the wall thickness.  Suter et al. in 2007 \cite{60} showed that the thermal shift rate of WGMs at a wavelength of 1550 nm in a silica microcapillary can be reduced when the capillary wall is 4 $\mu$m and the core is filled with water. This is because the refractive index of water has a negative dependence with increasing temperature and, since the optical modes of the hollow WGR have a portion of their fields in the water, it partially cancels out the positive dependence of the refractive index of the glass. For such resonators the equation used to describe the thermal shifting in hollow microcavities is given as \cite{60}
\begin{equation}
\frac{\triangle {\mathbf \lambda}}{{\mathbf \lambda}}=\alpha_{th} \triangle T+\frac{\partial n_{eff}}{\partial n_{wall}}\frac{{\kappa }_{wall}}{n_{eff}}\triangle T+\frac{\partial n_{eff}}{\partial n_{core}}\frac{{\kappa }_{core}}{n_{eff}}\triangle T,
\end{equation}

where $\alpha_{th}$ is the linear thermal expansion coefficient for glass, $n_{eff}$ is the effective refractive index of the mode, $\kappa_{core}$ and $\kappa_{wall}$  are the fraction of EM field in the core and the wall, respectively, $n_{core}$ and $n_{wall}$ are the refractive indices of the core and the wall material, respectively, and $\Delta T$ is a change in  temperature.   It was predicted in the same paper that the thermal shifting could be almost eliminated if the wall thickness were further reduced to 1.7 $\mu$m.
Not only can a negative thermo-optic coefficient material in the core reduce the red shift of the WGMs,  it can actually cause a significant blue shift of the higher order radial modes.  Recently, an ethanol-filled microbubble WGR was shown to have modes that experienced blue shifts at a rate of 100 GHz/K while other modes remained relatively stable \cite{42}. This is shown in the result of an FEM simulation plotted in Fig. 7(a) for an ethanol-filled microbubble.

\begin{figure}
%\resizebox{.5\columnwidth}{!}
{
\includegraphics[origin=r,scale=.2]{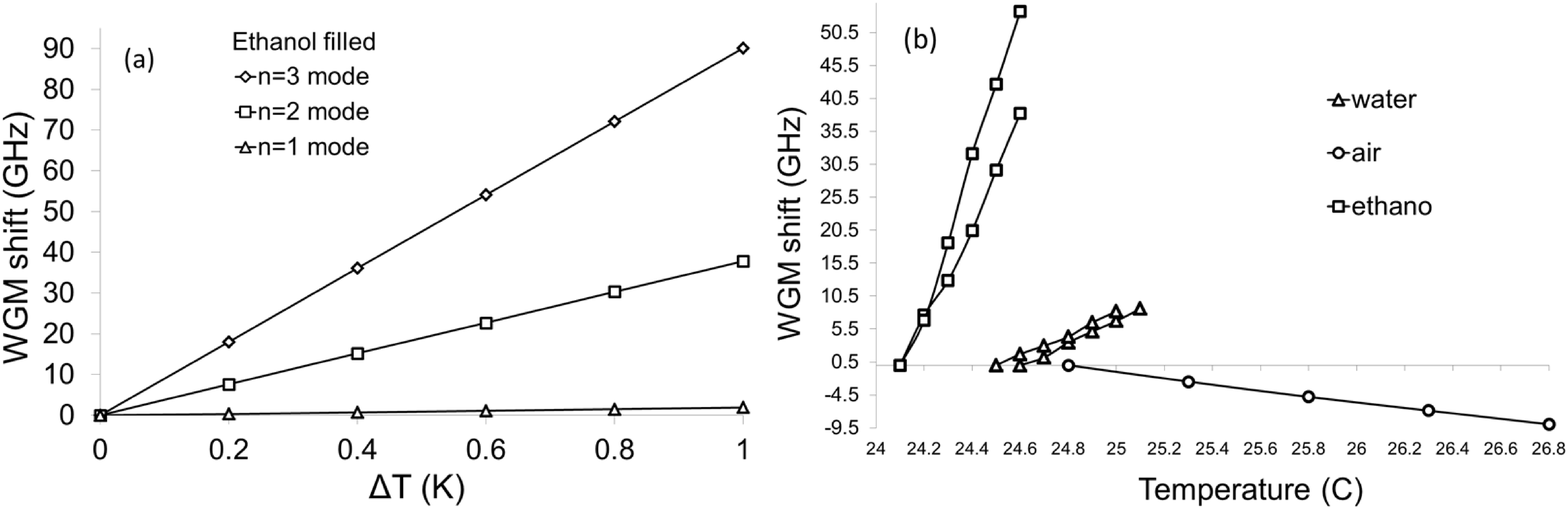} }
\caption{FEM simulation of the thermal shift for WGMs in a liquid-filled microbubble. (a) Ethanol-filled microbubble. The outer diameter is 60 $\mu$m and the wall thickness is 1 $\mu$m. (b) The experimental shift  for a water-filled, then air-filled and then ethanol-filled bubble. These results were taken in the same bubble at different times and different initial temperatures. The air-filled bubble is shifted over a larger temperature range.}       % Give a unique label
\end{figure}

The first order radial mode, $n = 1$, is confined mostly within the glass wall of the bubble; however, as the radial order is increased more of the mode enters the fluid core thus reducing its effective index. The thermo-optic coefficient of silica glass is $dn/dT$$=$$11\times$$10^{-6}\:/$K.  Thus, when an air-filled bubble is externally heated the optical modes red shift at a rate of around $-3$ GHz/K, similar to that obtained for a solid microsphere. However, $dn/dT$ for water is $-91\times$$10^{-6}\: /$K \cite{42}; therefore, when a water-filled bubble is heated the modes will shift in different directions at different rates depending on the radial order and wall thickness. When the bubble is filled with ethanol, which has a refractive index close to water,  $dn/dT$$=$$400\times$$10^{-6}\: /$K \cite{42}.   The resulting shift rates for increasing temperature for ethanol are plotted in Fig. 7(a). All the modes are blue shifted, but the $n = 1$ mode is only slightly shifted, while the $n=2$ and $n=3$ experience very large shift rates of 35 GHz/K and 90 GHz/K, respectively. The experimental results for a water-filled, air-filled and then ethanol-filled bubble are shown in Fig. 7(b). These results were taken in the same bubble at different times and for different initial temperatures. The air-filled bubble is shifted over a larger temperature range. This result highlights the fact that, when the wall thickness is on the order of the WGM wavelength, the mode shifting spectrum is actually quite complex.

\section{Refractive Index Sensing}
As we have seen, the WGMs in a hollow WGR are particularly sensitive to changes in the refractive index of the core material due to the significant overlap of the modes with the core. The first refractive index sensor using a hollow core  WGM resonator was demonstrated in 2004 by Moon et al. \cite{61} where the inside of a thick walled glass capillary was coated with a layer of dye-doped polymer. A frequency doubled, Q-switched Nd:YAG pulse laser at a wavelength of 532 nm was focussed on to the capillary as a pump source for the dye, and lasing emission from the dye was collected in the scattered field. The core of the capillary was filled with ethanol, the refractive index of which was changed by adding methanol. A refractive index change of 0.03 resulted in a 1 nm blue shift of the  WGMs with a lasing \emph{Q} factor around $10^6$.
In an experiment by White et al. in 2006 \cite{34} an LCOR was formed by stretching the centre section of a fused silica capillary under an H$_2$O flame until the outer radius reached around 40 $\mu$m. After pulling, the capillary was etched with low concentrations of HF acid to achieve the desired wall thickness. The resulting LCOR had an effective wall thickness of 3 $\mu$m. The capillary was filled with a solution of water and ethanol and the refractive index was changed by varying the concentration of the solution. The sensitivities obtained were approximately 0.56 nm/RIU and 2.6 nm/RIU (refractive index unit) for LCORs with wall thicknesses of 3.6 $\mu$m and 3 $\mu$m, respectively. The LCOR could theoretically detect a refractive index change of $1.8\times10^{-5}$ RIU.
This work was followed by an LCOR coupled to an anti-reflective ridge optical waveguide (ARROW) where the liquid filled capillary was evanescently coupled to a ridge waveguide on a chip \cite{62}. Shortly after this, a paper appeared comparing the refractive index sensing capability of a solid microsphere in a microfluidic channel to an LCOR in the form of a capillary with an effective wall thickness of 4 $\mu$m \cite{63}.
Around the same time another group \cite{64,65} also used thin walled capillaries as refractive index sensors. The thin walls allowed for refractive index sensitivity as large as 390 nm/RIU operating at a wavelength of 1500 nm in cores filled with microscope immersion oils. The overall resolution of the devices, however, was compromised by the poor \emph{Q} factor of around 500.
A fibre-coupled polymer tube formed in a polymer block was also used as highly sensitive refractive index sensor \cite{66}. A glass capillary LCOR was immersed into a curable low index polymer, which was later solidified by UV curing. After the capillary was embedded into the solid polymer matrix, the capillary wall was etched down to submicron values and eventually was fully eliminated, see Fig. 8. Therefore, a significant portion of the resonant WGMs was located inside the tested liquid (rather than propagating mainly along the capillary wall, as in other LCORs) and this considerably increases its sensitivity.

\begin{figure}
%\resizebox{.5\columnwidth}{!}
{
\includegraphics[origin=r,scale=.35]{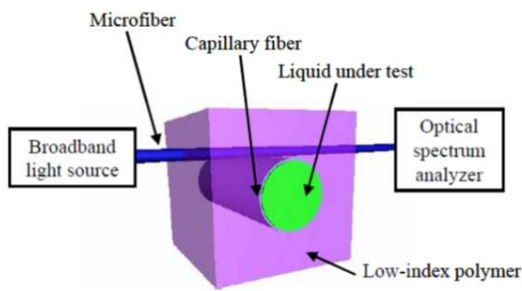} }
\caption{Diagram of a polymer encased liquid-core optical ring-resonator sensor. Reproduced from \cite{66} with permissions.}       % Give a unique label
\end{figure}

On-chip, self-assembled, rolled-up microtubes were used to explore refractive index sensing. The structure consists of a tube with 2–3 $\mu$m diameter, made out of one/two Si/SiO$_x$ layer rotations \cite{67}. The tube length was of the order of 100 $\mu$m$- 1$ mm. The microtube was filled with an aqueous solution of sugar by placing one open extremity of the tube in contact with a droplet deposited onto the substrate by the glass capillary. The optical response was studied by micro photoluminescence spectroscopy.  The achieved sensitivity was around 62 nm/RIU and the advantage of this device is the on-chip fabrication technique and the ability to make a large number of tubes from a wide variety of materials. Later improvement gave 450 nm/RIU \cite{49}.
An experimental study in \cite{68} showed that an LCOR in the form of a liquid-filled capillary could achieve an RI detection sensitivity of 570 nm/RIU with a \emph{Q} factor of $1.2\times10^5$ if higher order modes are used. A change of $2.8\times10^{-7}$ RIU was observed with a noise equivalent detection limit  of $3.8\times10^{-8}$ RIU. The operating wavelength was 980 nm and the wall thickness was about 1 $\mu$m for the bulk RI detection and 2 $\mu$m for the surface mass detection experiments.

LCORs tend to have thin walls, 5 $\mu$m or less, so that the evanescent field of the WGM in the wall of the capillary can extend into the core material.  However, it is also possible to measure the change in the refractive index of the core material in thick wall capillaries by coating the inner wall with a fluorescent material such as quantum dots (QD). The fluorescent material can be optically pumped through the capillary wall using a focussed laser or an LED light source as demonstrated by \cite{69}. The fluorescence from the QDs was coupled into the capillary wall where it was trapped as WGMs. The WGMs were observed in the scattered spectrum of the QD emission and the peaks of the WGM shifted as a function of the refractive index of the core material. Capillaries with wall thickness of 135 $\mu$m and 30 $\mu$m showed a refractometric sensitivity of 9.8 nm/RIU and 6.8 nm/RIU, respectively, and a maximum detection limit of $\sim7.2\times10^{-3}$ RIU.  A similar approach was followed in 2013 with a fluorescent polymer coating inside a thick walled capillary, achieving a sensitivity of 30 nm/RIU \cite{70}.
All the LCOR refractive index sensors discussed so far were made from straight capillaries. In 2011 a high \emph{Q} microbubble WGR refractive index sensor was studied \cite{43}. The bubble was filled with ethanol and WGM positions were recorded for increasing concentrations of water. This device had a limit of detection of $10^{-6}$ RIU. However, the sensitivity was low at 0.4 nm/RIU due to the much thicker bubble wall which was around 4 $\mu$m. The operating wavelength was 1550 nm.

\section{LCOR WGM for Biosensing}
Perhaps the most interesting or maybe even the most suitable application for hollow WGRs is in the area of biosensing. Biosensing using WGMs in solid WGRs has been studied extensively both experimentally and theoretically over the last two decades.  In the case of the solid WGR, the resonator is immersed in the fluid containing the molecular species of interest. This setup creates problems with regards to coupling to an optical waveguide and the rapid exchange of the sample fluid. Also, in the solid WGR, only a small percentage of the mode interacts with the fluid through the evanescent field. Despite these drawbacks solid WGRs have shown remarkable detection capabilities down to the single virus \cite{71}, single nanoparticle \cite{72}, and single molecule level \cite{73}.
The hollow WGR can overcome some of the problems associated with the solid WGR and is probably a more realistic and robust device for real world applications. The fluid containing the sample material can be flowed continuously and changed rapidly without disturbing the waveguide coupler. When the hollow WGR is operated in the quasi-droplet regime there is maximum overlap of the mode with the fluid and this dramatically increases the sensitivity.
The mechanisms relating to biosensing and nanoparticle detection were already discussed in the sections above. Here, we will give an overview of the state-of-the-art. One of the first examples of biosensing using a hollow WGR was demonstrated by White et al. in 2006 \cite{62} where the inside of a hollow glass capillary was activated by coating with 1\% (v/v) 3- minopropyltrimethoxysilane in 10/90 (v/v) distilled water/ethanol which was passed through the hollow capillary. Detection of biomolecules was performed when 0.05 mg/ml of Bovin Serum Albumin (BSA) in a phosphate buffer was sent through the capillary for approximately 3 minutes while monitoring the WGM spectral position. The WGMs were excited by the ARROW waveguide system described earlier; the setup is shown schematically in Fig. 9 \cite{62}.

\begin{figure}
%\resizebox{.5\columnwidth}{!}
{
\includegraphics[origin=r,scale=.2]{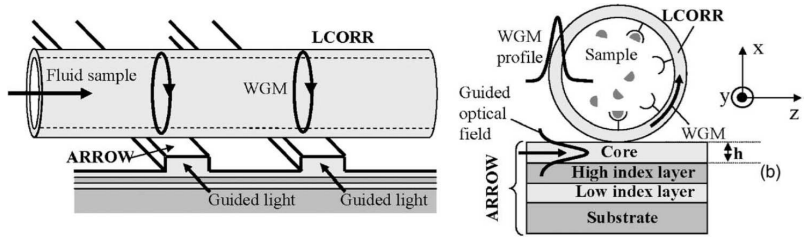} }
\caption{Antiresonant reflecting optical waveguide ARROW structure coupled to a liquid core optical ring resonator biosensor. Reproduced from  \cite{62} with permissions.}       % Give a unique label
\end{figure}
Since this work, numerous experimental papers have been published showing the biosensing capability of the WGM LCOR and here we highlight just a few of them.  In 2008, detection of streptavidin was achieved using a phage that was coated to the inside of a glass capillary \cite{74}. The filamentous phage R5C2 has peptides that bind specifically to streptavidin. The phage based WGM LCOR has some advantages over the antibody type optofluidic biosensor because the phage is more robust, plus it is cheaper and faster to implement. The phage-based WGM LCOR was able to achieve an experimental detection limit of 100 nM and theoretical limit of 5 nM with some room for improvement by improving \emph{Q} and surface chemistry. The detection limit is comparable to other detection systems such as surface plasmon resonance and quartz crystal microbalance.
Also in 2008, the same group used a WGM LCOR for the label free quantitative detection of DNA \cite{75}. In this setup a temperature stabilised WGM LCOR with a diameter of 100 $\mu$m and wall thickness of 4 $\mu$m having a bulk refractive index sensitivity (BRIS) of 37 nm/RIU was used to detect the presence of various DNA samples that had different strand lengths (25-100 bases), number of base mismatches (1-5) and concentrations (10 pM to 10 M).  It was shown that the LCOR was sensitive enough to differentiate DNA with only a few base-mismatches based on the raw sensing signal and kinetic analysis, i.e. the time response of the WGM due to different DNA combinations on the wall of the LCOR. Bulk DNA detection of a 10 pM 25-mer was achieved with a mass loading limit of detection estimated to be 4 pg/mm$^2$. The authors claimed that there was room for improvement by reducing the capillary wall thickness, thus increasing the BRIS.   Following this work, in 2009 Zhu et al. looked at  label free detection of the breast cancer marker, CA15-3 \cite{76}.  It was shown that the WGM optofluidic system was capable of detecting CA15-3 in human serum. Shortly after this the WGM LCOR was used to detect the presence of another breast cancer marker, this time HER2  \cite{77}. Other biomarkers detected using the WGM LCOR were CD4+ and CD8+ T-Lymphocyte whole cells and CD4+ T-Lymphocyte cell lysis which are indicators of HIV \cite{78}. Also, in 2010, label-free DNA methylation analysis using optofluidic ring resonators was achieved \cite{79}. Further examples include detection of pesticides \cite{80} and glucose \cite{81}.
One of the main problems with all types of WGM sensors is thermal drift due to thermal expansion and thermo refractive noise. These result in a common mode noise that affects all the optical modes of the resonator. The common mode noise puts a limit on the detection capability of WGM sensors.  One way to virtually eliminate this noise is to use  self-referencing. In WGRs, this relies on the splitting of the cavity mode due to backscattering. This technique was demonstrated in a microtoroid WGR for the detection of single nanoparticles \cite{72}.  However, for the detection of multiple particles the backscattered splitting effect is not desirable. Instead, mode splitting can be achieved by using co-resonance in coupled cavities. Recently, mode splitting in coupled microbubble resonators was used to increase the sensitivity  of biomolecular detection \cite{82}. The coupled microbubble system is shown in Fig. 10. Detection of binding of biomolecules as low as 2 pg/mL was achieved with a noise equivalent detection limit (NEDL) of only 0.5 pg/mL. These results make the self-referenced, optofluidic ring resonator (SR-OFRR) based label-free biosensor comparable with the widely used enzyme-linked immunosorbent assay-based.

\begin{figure}
%\resizebox{.5\columnwidth}{!}
{
\includegraphics[origin=center,scale=.3]{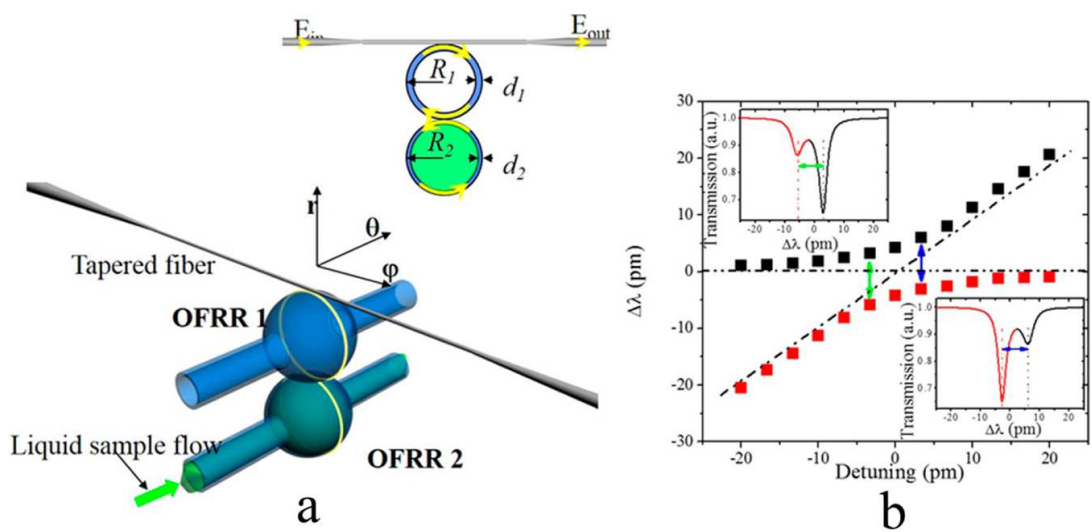} }
\caption{(a) Schematic of a self-referenced, optofluidic ring resonator (SR-OFRR) biosensor.  OFRR 1 acts as the reference resonator, while the liquid-filled OFRR 2 serves as the sensing resonator. (b) Theoretical mode splitting in an SR-OFRR biosensor with resonance detuning. The red and black squares represent the left and right branches of the modes, respectively. Reproduced from  \cite{82} with permissions. }       % Give a unique label
\end{figure}

Another self-referencing method was  achieved recently in a single resonator system where one mode was thermally compensated by using three different modes:  a sensing mode and two reference modes \cite{83}. Both of the reference modes are insensitive to the adsorption of the target molecules, but have the same thermal and mechanical noise levels as the sensing mode. The noise from the reference modes is subtracted from the signal mode, thereby eliminating the thermal drift noise and mechanical noise.

\section{Biosensing with Liquid Core Lasers}
WGRs have proven to be efficient laser cavities where the WGR simultaneously forms the gain medium and resonator, thus they could be described as mirrorless microlasers.  Recently, hollow microcavities have also been employed as laser resonators. In this case, the hollow core of the resonator is filled with a gain medium such as a laser dye. The first dye-filled capillary laser was demonstrated in 1992 \cite{84}, but these devices had thick capillary walls (tens to hundreds of microns) and required dyes with refractive indices higher than the glass such that the WGM was circulating within the liquid. Similar examples followed shortly afterwards \cite{61,85} and these used slightly protracted methods that did not allow the high \emph{Q} modes of the glass capillary to directly interact with the dye. In 2007, lasing using a low refractive index was achieved in a hollow microcapillary \cite{86} with thin walls of a few micron. The wall of the LCOR was thin enough to allow the high \emph{Q} WGMs to be exposed to the core to support the low-threshold laser oscillation. The WGM had the evanescent field on the outer surface thus enabling out coupling of the laser emission through optical waveguides in contact with the surface. A typical dye laser setup is shown in Fig. 11(a).  On-chip dye lasers using integrated microfluidic ring resonators have also been demonstrated \cite{90}.

\begin{figure}
%\resizebox{.5\columnwidth}{!}
{
\includegraphics[origin=r,scale=.2]{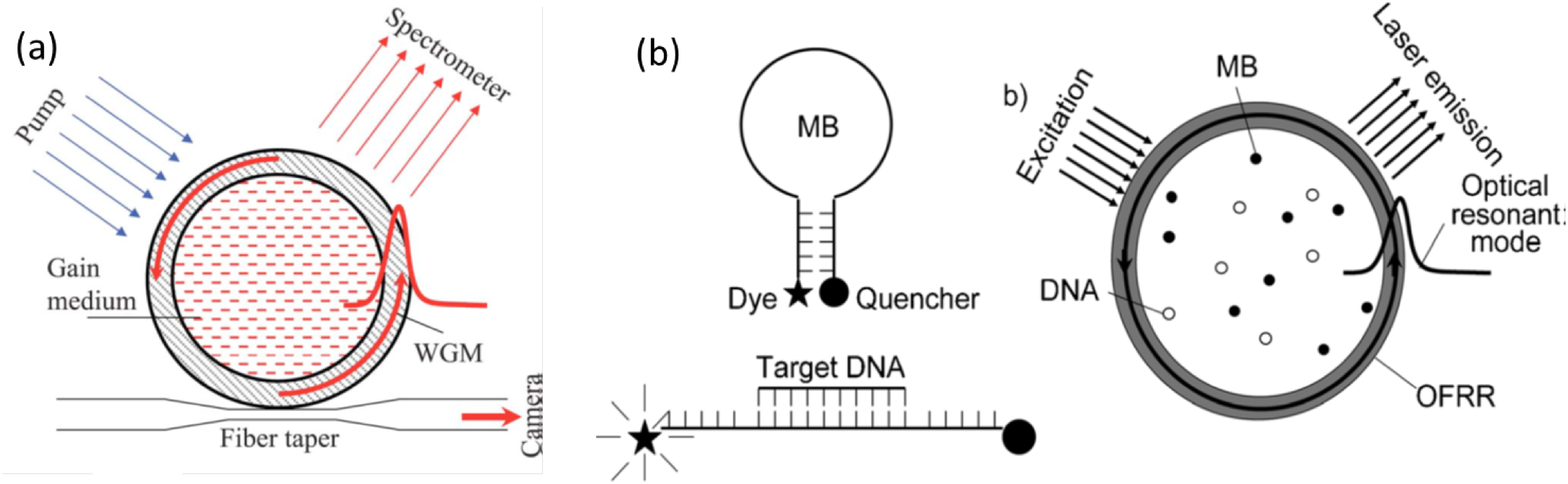} }
\caption{(a) Diagram of a thin-walled WGM LCOR dye laser system coupled to a tapered optical fibre. Reproduced from  \cite{86} with permissions. (b) Schematic of a molecular beacon dye laser used for the detection of mismatched DNA strands. Reproduced from \cite{106} with permissions.}
       % Give a unique label
\end{figure}

Liquid core WGM lasers can have applications in biosensing; recently a molecular beacon probe was used as a gain medium in an optofluidic ring resonator \cite{92}. The beacon consists of a DNA strand with a dye molecule and quencher on opposite ends. The beacon can be in an open or closed configuration, see Fig. 11(b), when it is open low threshold lasing emission is possible.  When the beacon is closed the dye molecule emission is quenched and spontaneous fluorescence is observed. When the target DNA is perfectly matched to the beacon it is in the open configuration; if the target DNA has a single base mismatch then the beacon is closed. The small difference in binding affinity between the perfectly matched DNA and the single mismatch DNA causes a small change in the laser gain coefficient. This small change amplifies laser emission intensity by orders of magnitude because of the strong optical feedback that is provided by the optical cavity.

Also in 2012, a WGM LCOR was used for a DNA matching technique called high resolution melting (HRM). The sensing signal of HRM uses fluorescence from dyes that have been inserted into the DNA \cite{93}.  When the dye is bound to double-stranded DNA it has strong fluorescence. With increased temperature, double-stranded DNA dissociate into single-stranded DNAs. As a result, the dye is released from the double-stranded DNA and its fluorescence decreases. The DNA melting curve was acquired by monitoring the fluorescence as a function of temperature. The target and the base-mismatched DNA can be discriminated by analysing corresponding melting curves.  In contrast to molecular beacons and DNA arrays HRM is capable of analysing DNA sequences of hundreds of bases. In a similar vein, the variation of the lasing threshold and lasing intensity of peptide bound proteins in a WGM LCOR was used to monitor and to quantify protein-protein interactions \cite{94}.

\section{Gas Sensing}
Hollow core optical waveguides have been used for the optical detection of gases and vapours. The hollow microcapillary WGRs were internally coated with a polymer whose refractive index changed when exposed to a solvent. The refractive index change resulted in a shift of the WGMs, the amount and rate of the shifting is a signature of the particular solvent. A schematic of the so-called capillary-based, optical ring resonator gas chromatograph is shown in Fig. 12 \cite{95}.

\begin{figure}
%\resizebox{.5\columnwidth}{!}
{
\includegraphics[origin=r,scale=.4]{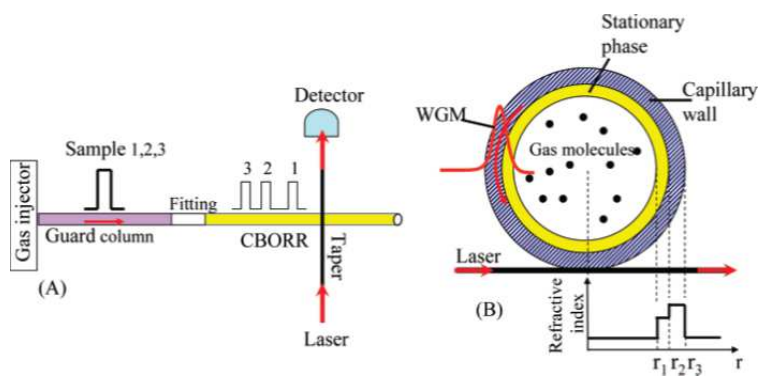} }
\caption{Setup for a capillary-based, optical ring resonator gas chromatograph. (A) The sample is introduced in the gas injector, after which it travels as a short pulse carried by high purity hydrogen gas. The separation of different compounds in the sample is enabled by their different interaction with the stationary phase. Optical detection is performed by measuring the transmission through a tapered optical fibre in contact with the capillary. (B) Cross- section of the capillary based optical ring resonator gas chromatograph showing the WGM confined in the capillary wall. Reproduced from \cite{95} with permissions.}       % Give a unique label
\end{figure}

In the first example of its kind, a 200 nm thick coating of Polyethylene Glycol (CarboWax) was deposited onto the inner wall of the capillary. The CarboWax is a substance that retains polar molecules. Gas samples of ethanol, toluene, decane, and dimethyl methylphosphonate (DMMP) were extracted from  saturated vapour in the head space of sample vials using a solid-phase microextractor. The selection of analytes covers polar and nonpolar organic compounds. From the extractor the vapours passed into a gas chromatograph injection system, heated to 250\celsius, which generated a gas pulse. The gas pulse travels along a 1.8 m passive fused silica guard column to the capillary-based, optical ring resonator (CBORR)  gas chromatograph. The WGMs were excited using a tapered fibre and a laser operating at 1550 nm. The WGM shift reflected the interaction between the gas analyte and the CarboWax.   In the following year, the same technique was used for the rapid detection of chemical vapours such as ethanol and hexane, but using different stationary phases such as OV-17 (which contains 50\% polar group and 50\% nonpolar groups) and PEG-400 which is highly polar \cite{96}. Similarly DNT, an explosive compound, was detected using a PEG-1000 stationary phase \cite{97}. The detection limit of the vapour sensor for DNT was approximately 200-300 pg for a WGM resolution of 0.15 pm.  However, the device sensitivity was shown to be quite dependent on the wall thickness, ranging from 0.41 pm/ng to 0.67 pm/ng for two devices made by the same fabrication process.   Further improvements were made to the system by adding a second tapered fibre on the sensing capillary. One taper was placed on or  near the entrance and the second taper was placed a few centimetres downstream, thus enabling multi-point, on-column, real-time detection of a number of vapour molecules. The separation and detection of twelve analytes with various volatilities and polarities was demonstrated within four minutes. Further analysis of the hollow core ring resonator gas detection system was presented in \cite{98}.

\section{Conclusion}
An overview of the sensing ability and sensing applications of hollow/liquid core whispering gallery resonators has been given. Although not every paper in the field was discussed we attempted to cover the most important results since the inception of the first LCOR device. It should be evident to the reader that  hollow core WGRs are unique among optical sensors. High \emph{Q} LCORs achieve all the same sensing capabilities as the traditional solid WGRs, plus they have the additional features associated with microfluidic channel systems, making LCOR devices more suitable for real-world applications than solid WGRs. As mentioned, there are many topics that were not discussed in this review; however, the relatively new field of microcavity optomechanics is one area that deserves a comment in our final remarks. Cavity optomechanics is the study of how minute optical forces affect the internal degrees of motion in micron-scale, dielectric structures. Due to the interaction between the intracavity light field and the mechanical motion in such high \emph{Q} resonators the photon pressure can damp or enhance the material's mechanical motion. This effect has led to many interesting physical phenomena and put WGRs at the forefront in the study of the boundary between the quantum and classical world. Recently, practical sensing applications have been demonstrated using microcavity mechanical modes as the measured parameter. The thin wall of the LCOR allows the mechanical modes to interact with any fluid in the core such that a change in the fluid results in a change in the mechanical modes' frequency. Using this principle, the mechanical modes of an LCOR were used to measure the viscosity of sucrose solutions ~\cite{102,103,104} and may even be used for the detection of bio/nano particles \cite{105}.  Given these latest results, it is clear that the sensing capabilities of hollow/liquid core whispering gallery resonators is continuing to evolve at a rapid pace.

%\begin{equation}
%\frac{\triangle \omega }{\omega }=-\frac{{\alpha }_{ex}{\sigma }_p}{{\varepsilon }_0R\left(n^2_s-n^2_m\right)}
%\end{equation}
%\begin{equation}
%\frac{\triangle \lambda (p_i)}{\lambda }=\frac{2n_0b^3+12CGb^3}{4Gn_0(a^3-b^3)}p_i-\frac{n_0(a^3+b^3)+12CGa^3}{4Gn_0(a^3-b^3)}p_i
%\end{equation}
%\begin{equation}
%\frac{\triangle {\mathbf \lambda }}{{\mathbf \lambda }}=\alpha \triangle T+\frac{\partial n_{eff}}{\partial n_{wall}}\frac{{\kappa }_{wall}}{n_{eff}}\triangle T+\frac{\partial n_{eff}}{\partial n_{core}}\frac{{\kappa }_{core}}{n_{eff}}\triangle T
%\end{equation}
%\begin{figure}
%% Use the relevant command for your figure-insertion program
%% to insert the figure file.
%% For example, with the option graphics use
%%\resizebox{0.75\columnwidth}{!}{%
%%  \includegraphics{fig1.eps} }
%\caption{Please write your figure caption here.}
%\end{figure}
%\begin{table}
%\caption{Please write your table caption here.}
%\begin{tabular}{lll}
%\hline\noalign{\smallskip}
%first and second and third  \\
%number and number and number \\
%number and number and number \\
%\end{tabular}
%\end{table}
%

\end{document}